\documentclass[conference]{IEEEtran}
\IEEEoverridecommandlockouts
\usepackage{amsmath}
\usepackage{soul}
\usepackage{amsfonts}
\usepackage{amssymb}
\usepackage{amsthm}
\usepackage{abraces}
\usepackage{romannum}
\usepackage{multirow}
\usepackage{multicol}
\usepackage{longtable}
\usepackage{array}
\usepackage{bm}
\usepackage{tikz}
\usepackage{hyperref}
\usepackage{cleveref}

\crefname{lemma}{Lemma}{Lemmas}
\usepackage{cite}
\usepackage{adjustbox}
\usepackage{pgfplots}
\pgfplotsset{compat=1.18}
\usetikzlibrary{shapes.geometric,arrows,backgrounds,fit,positioning,fadings,through,patterns}
\usepackage{verbatim}
\usepackage{algpseudocode,algorithm,algorithmicx}
\usepackage{setspace}
\usepackage{graphicx}
\usepackage{caption}
\usepackage{subcaption}
\usepackage{stfloats}
\usepackage{siunitx}
\usepackage{fixltx2e}
\usepackage{psfrag}
\usepackage{color}
\tikzset{>=latex'}
\usetikzlibrary{calc}
\usepackage{orcidlink}

\makeatletter
\renewenvironment{thebibliography}[1]{%
\@xp\section\@xp*\@xp{\refname}%
\normalfont\footnotesize\labelsep .5em\relax
\renewcommand\theenumiv{\arabic{enumiv}}\let\p@enumiv\@empty
\vspace*{-2pt}% NEW
\list{\@biblabel{\theenumiv}}{\settowidth\labelwidth{\@biblabel{#1}}%
\leftmargin\labelwidth \advance\leftmargin\labelsep
\usecounter{enumiv}}%
\sloppy \clubpenalty\@M \widowpenalty\clubpenalty
\sfcode`.=\@m
}{%
\def\@noitemerr{\@latex@warning{Empty `thebibliography' environment}}%
\endlist
}
\makeatother

\let\OLDthebibliography\thebibliography
\renewcommand\thebibliography[1]{
\OLDthebibliography{#1}
\setlength{\parskip}{-0.5pt}
\setlength{\itemsep}{-0.1pt}
}

\title{Covertness in the Near Field:\\ Maximizing the Covert Region with FDA 
\thanks{This work was funded by the Federal Ministry of Education and Research of the Federal Republic of Germany under grant number 16KISK037 (6GEM).}} 
\author{\IEEEauthorblockN{Fatemeh Lotfi\IEEEauthorrefmark{1}\,\orcidlink{0000-0001-7102-3216}, Stefan Roth\IEEEauthorrefmark{1}\,\orcidlink{0000-0003-4408-3306}, Anas Chaaban\IEEEauthorrefmark{3}\,\orcidlink{0000-0002-8713-5084}, and Aydin Sezgin\IEEEauthorrefmark{1}\,\orcidlink{0000-0003-3511-2662}}
\IEEEauthorblockA{\IEEEauthorrefmark{1}Ruhr University Bochum, Bochum, Germany\\
\IEEEauthorrefmark{3}University of British Columbia, Kelowna, Canada\\
Email: \{fatemeh.lotfi, stefan.roth-k21, aydin.sezgin\}@rub.de, anas.chaaban@ubc.ca}}

\begin{document}
\maketitle

\begin{abstract}
    Covert communication in wireless networks ensures that transmissions remain undetectable to adversaries, making it a potential enabler for privacy and security in sensitive applications. However, to meet the high performance and connectivity demands of sixth-generation (6G) networks, future wireless systems will require larger antenna arrays, higher operating frequencies, and advanced antenna architectures. This shift changes the propagation model from far-field planar-wave to near-field spherical-wave which necessitates a redesign of existing covert communication systems. Unlike far-field beamforming, which relies only on direction, near-field beamforming depends on both distance and direction, providing additional degrees of freedom for system design. In this paper, we aim to utilize those freedoms by proposing near-field Frequency Diverse Array (FDA)-based transmission strategies that manipulate the beampattern in both distance and angle, thereby establishing a non-covert region around the legitimate user. Our approach takes advantage of near-field properties and FDA technology to significantly reduce the area vulnerable to detection by adversaries while maintaining covert communication with the legitimate receiver. Numerical simulations show that our methods outperform conventional phased arrays by shrinking the non-covert region and allowing the covert region to expand as the number of antennas increases. 
\end{abstract}

\section{Introduction}
    To fulfill the increasing demand for high-speed and high-capacity communication, the upcoming sixth-generation (6G) wireless networks promise to integrate innovative technologies such as extremely large-scale antenna arrays and additional frequency bands. 
    These advancements, however, would lead to significantly different electromagnetic properties, shifting from planar-wave to spherical-wave propagation and resulting in unavoidable near-field effects \cite{liu2023near}. The spherical-wave propagation model, unlike traditional planar-wave models, captures both direction and distance information of the receiver. This precise characterization enables \textit{beamfocusing}, allowing array radiation patterns to concentrate on specific spatial points. Hence, near-field modeling is required to include the case that the distance between the transmitter and receiver can be less than or equal to the Rayleigh distance (see \cite{liu2023nearcomm}). 

    Meanwhile, the open nature of wireless channels leaves communication signals vulnerable to detection and interception by third parties, posing risks of revealing communication intentions or the leakage of sensitive information. In this regard, physical layer security (PLS), as a complement to encryption techniques, is proposed to safeguard private information from eavesdropping. Nonetheless, in some scenarios, even the detection of communication activity can cause privacy disclosure, and it is necessary to transmit messages covertly. Covert communication addresses this need by enabling a transmitter to send information to an intended receiver while remaining undetectable by a warden, thus providing an effective solution for privacy-sensitive applications.    
    
    In \cite{bash2013limits}, the fundamental limits of covert communications are analyzed. Notably, multi-antenna systems with full knowledge of the warden's channel can facilitate covert transmission by forcing the warden into the null space of the beampattern while directing the transmit signal toward the covert receiver \cite{chen2021multi, lotfi2023under}. With less available channel information, subsequent studies show that covert communication is still achievable when the adversary faces uncertainties due to noise \cite{he2017covert}, channel fading \cite{shahzad2017covert}, full-duplex devices \cite{shahzad2018achieving}, or interference \cite{soltani2018covert}. However, satisfactory covert performance may not be achievable if the channel between the transmitter and warden aligns with the channel between the transmitter and the desired receiver (see, e.g., \cite{li2024enhancingsecrecyratedirectionrange}), as traditional techniques have limitations in effectively differentiating these channels. 
    Therefore, in \cite{ma2022covert}, a promising beamforming scheme based on a frequency diverse array (FDA) is used, which allocates different frequencies to the antenna elements on the transmitter side to create a unique channel between the legitimate parties. 
    The study \cite{zhang2023distance} then considers enhancing the covert rate by jointly optimizing both the transmit beamforming vector and the FDA-specific frequency increments. 
    The studies \cite{li2023covert, cheng2024fda} further explore the effectiveness of FDA techniques in enhancing covert communication performance by employing a random FDA approach and FDA-aided covert communication as a strategy in multi-warden scenarios, respectively.  
    However, most of the existing studies on covert communication do not cover the near-field effects, even though both FDA and near-field communication offer potential benefits for covertness by enabling spatial focusing of signal power toward the intended receiver. 
    
    In \cite{zhang2023robust}, FDA-aided near-field beamfocusing was introduced to fill this gap by generating a highly focused beampattern that also decorrelates the communication and detection channels, thereby enhancing covert communication performance. 
    In \cite{hu2024minimizing}, the concept of a vulnerable region surrounding the legitimate user is introduced, which needs protection from the warden’s presence, and covert communication is maintained as long as the warden remains outside this defined vulnerable region. 
    Yet, to the best of our knowledge, the existing literature on near-field covert communication does not address how system parameters influence the size of the covert region.  
    
    \paragraph*{Contributions} 
    The goal of this study is to achieve a strong energy-focusing effect for improved communication between legitimate devices, such that the energy leakage toward potential wardens located outside the intended user's non-covert region is reduced. 
    By considering the distance-dependent spherical wavefronts, we exploit near-field channel characteristics to enhance the system's covertness. This is achieved by employing FDA-based transmission techniques while allowing the warden to move freely inside the covert region. In particular, we propose to employ linear, random, and optimized frequency offset designs at the transmitter to manipulate distance- and angle-dependent beampatterns and reduce the non-covert region. 
    \paragraph*{Notation}Scalars and vectors are denoted in italic and bold italic letters, respectively. $\|\cdot\|$ indicates the $L_2$-norm, while $\mathbb{E}[\cdot]$ refers to the expectation and $Q^{-1}(.)$ is the inverse Q-function. $\circ$ denotes Hadamard product and the notation $\lfloor.\rfloor$ indicates the floor function. 
    
\section{System Model}
    We consider a covert communication scenario, where a base station (Alice) aims to covertly transmit data to a user device (Bob) while being monitored by a warden (Willie). Thereby, Bob and Willie are located in the near-field of Alice, in which Willie can roam freely as long as not entering the covert region to be specified later. Alice is equipped with an $N$-element antenna array and uses different frequencies across its array elements, with $f_n$ indicating the frequency selected for transmission at antenna $n$. Bob and Willie are equipped with a single antenna each. A two-dimensional cell is considered, where Alice's antennas are aligned along the y-axis, with the first antenna element at the origin and an inter-element spacing of $d=\lambda_c/2$, in which $\lambda_c$ denotes the carrier wavelength.  
    Utilizing the polar coordinate system, the coordinate of any receiving point in space is given by $(r_i, \theta_i)$, where the subscript $i\in \{b, w\}$ denotes Bob or Willie.
    Within the described setup, the boundary between the near-field and far-field regions is determined by the Rayleigh distance $\frac{2D^2}{\lambda_c}$ \cite{liu2023nearcomm}, where $D=(N-1)d$ represents the aperture of the antenna array. This implies that the distances from Alice to Bob and Willie are not exceeding $\frac{2D^2}{\lambda_c}$.  
 
    \subsection{Near-Field Channel Model}
    The near-field channel model is obtained by considering spherical wavefront. Due to this, the channel between Alice's $n$-th antenna and each of the two receivers $i$, located at $(r_i, \theta_i)$, is given by 
    \begin{align}
        h_{i, n} (f_n) = \beta(r_i) \exp{\left(-j\frac{2 \pi f_n}{c}(r_n(r_i, \theta_i) - r_i)\right)},
    \end{align}
    where $\beta(r_i) =\frac{\lambda_c}{4\pi r_i}$ denotes the channel gain, and $r_n(r_i,\theta_i)$ is the distance from the $n$-th antenna to receiver $i$, approximated using the Fresnel approximation as \cite{selvan2017fraunhofer} 
    \begin{align}\label{FresnelApp}
        r_n (r_i, \theta_i) \approx r_i + \frac{n^2 d^2 }{2r_i} - n d \sin{\theta_i} .
    \end{align}
    
    Under the assumption that the line-of-sight (LoS) components of all wireless channels are dominant, the channel vector between Alice and receiver $i$ can be expressed as 
    \begin{align}\label{Channels}
        \bm{h}_i (\bm{f}) & = \frac{1}{N}\left[h_{i, 1} (f_1), \dots, h_{i, N} (f_N)\right]^T, & i\in \{b, w\}.
    \end{align}
    Thereby, $\bm{f} = [f_1, \dots, f_n, \dots, f_N]$ represents the concatenation of all frequencies used at each antenna $n$. 
   In the considered system, we assume that Alice has precise knowledge of Bob's location and perfect information about the Alice-Bob channel due to the cooperation between legitimate nodes. However, Willie's location is not known to Alice, such that the worst-case of all channel realizations outside of the covert region is relevant. 

    \subsection{Signaling Model}
    We assume that Alice employs maximum ratio transmission (MRT) beamforming \cite{zhang2022beam}, which maximizes the rate between Alice and Bob, i.e., $\bm{w}(\bm{f}) = \frac{\bm{h}_b(\bm{f})}{\|\bm{h}_b(\bm{f})\|}$. 
    Thus, Alice's transmit signal can be written as 
    \begin{align}\label{Trans.Sig.}
        \bm{x}_a[l] = \sqrt{P_t} \bm{w}(\bm{f}) s[l],
    \end{align}
    where $P_t$ is Alice's transmit power, and $s[l]$ denotes the transmitted symbol with $l=1, \dots, L$ being the index of each channel use.  
    The information symbols are assumed to be independently distributed and have unit power, i.e., $\mathbb{E}[\left|s[l]\right|^{2}]=1$. 
    According to \eqref{Trans.Sig.}, the received signal at Bob is given by \footnote{It is important to note that this expression represents a simplified form of the received signal after processing through the FDA’s receive chain. As discussed in \cite{zhang2023distance}, first, a sequence of mixers is employed to eliminate the FDA's time-dependent components. The resulting signals are then filtered using low-pass filters and subsequently sampled by analog-to-digital converters. Finally, the $N$ processed streams are combined to produce the output.}
    \begin{align}
        y_b[l] = & \bm{h}_b^H(\bm{f})\, \bm{x}_a[l] + z_b[l], & l\in\{1,\dots,L\}, 
    \end{align}
    where $z_b \sim \mathcal{CN}(0,\sigma_{b}^2)$ is the additive white Gaussian noise (AWGN) observed at Bob which has a covariance of $\sigma_{b}^2$. 
    Thus, the received signal-to-noise ratio (SNR) observed by Bob for each channel use can be expressed as 
    \begin{equation}
        \gamma_{b} = \frac{P_t |\bm{h}_b^H (\bm{f})\, \bm{w}(\bm{f})|^2}{\sigma_{b}^2}. 
    \end{equation} 
    
    If the desired frame error probability at Bob is fixed to $\delta$, the achievable rate of the Alice-to-Bob channel can be approximated by the Polyansky-Poor-Verd{\'u} formula \cite{polyanskiy2010channel} as 
    \begin{equation}\label{CovertRate}
        R_{b} \approx \log_2 \left( 1 + \gamma_{b} \right) - \sqrt{\frac{\gamma_{b}(\gamma_{b}+2)}{L(\gamma_{b}+1)^2}} \frac{Q^{-1}(\delta)}{\ln{2}}.
    \end{equation}
    Although smaller $L$ lowers the achievable transmission rate, it also reduces Willie's chance of detecting the communication due to the limited number of observations.
    
    \subsection{Attacker Model}
    Willie attempts to understand whether Alice is transmitting data by performing an optimal statistical hypothesis test on the observed vector $\bm{y}_w=\{y_w[l]\}_{l=1}^{L}$ in a communication slot. To achieve this, Willie needs to distinguish between two hypotheses: the null hypothesis (${H}_0$), which corresponds to Alice not transmitting, and the alternative hypothesis (${H}_1$), which corresponds to Alice transmitting, i.e., 
    \begin{align} 
    y_{w}[l] =
        \begin{cases}\label{BHT@Warden}
         z_{w}[l]\, , & {H}_0,\\
         \sqrt{P_t}\, \bm{h}_w^H(\bm{f})\, \bm{w}(\bm{f}) s[l] + z_{w}[l]\, , & {H}_1, 
        \end{cases}
    \end{align} 
    where $\bm{h}_w(\bm{f})$ is Alice-Willie channel and $z_w \sim \mathcal{CN}(0,\sigma_{w}^2)$ is AWGN at Willie. Therefore, the average received signal energy per channel use observed by Willie is
    \begin{align}\label{Energy_W}
        \mathbb{E}\left[|y_{w}[l]|^2\right] = \begin{cases}\sigma_{w}^2& {H}_0,\\P_t\,|\bm{h}_w^H (\bm{f})\, \bm{h}_b (\bm{f})|^2 + \sigma_{w}^2& {H}_1.\end{cases}
    \end{align}
    To evaluate Willie's detection performance, the Kullback-Leibler (KL) divergence is adopted as the metric \cite{bash2013limits}, given by\looseness-1
    \begin{align}\label{KL_Div}
        {D}(P_0 \| P_1) \leq 2\epsilon^2.
    \end{align}
    Thereby, $P_0$ and $P_1$ denote the probability distributions of $\bm{y}_w$ under hypotheses ${H}_0$ and ${H}_1$, respectively, and $\epsilon$ represents the desired level of covertness, with smaller values indicating a higher level of covertness. 
    Based on \eqref{BHT@Warden}, $P_0$ and $P_1$ can be respectively derived as
    \begin{align*}
        P_0 
        & = \prod\limits_{l=1}^{L} f\left(y_{w}[l]|{H}_0\right) 
        = \frac{1}{\left(\pi\sigma_{w}^2\right)^L}\, e^{-\frac{\bm{y}_w^H\bm{y}_w}{\sigma_{w}^2}},\\
        P_1 & = \prod\limits_{l=1}^{L} f\left(y_{w}[l]|{H}_1\right) \nonumber\\
        & = \frac{1}{\pi^L\left(P_t |\bm{h}_w^H(\bm{f})\, \bm{w}(\bm{f})|^2+\sigma_{w}^2\right)^L}\, e^{-\frac{\bm{y}_w^H\bm{y}_w}{P_t |\bm{h}_w^H(\bm{f})\, \bm{w}(\bm{f})|^2+\sigma_{w}^2}}.
    \end{align*}
    Substituting the above expressions in \eqref{KL_Div}, KL divergence can be expressed as 
    \begin{align}
        {D}(P_0 \| P_1) 
        & = L \left[\nu - \ln{\left(1+\nu\right)} \right], 
    \end{align}
    where $\nu = \frac{P_t |\bm{h}_w^H(\bm{f})\, \bm{w}(\bm{f})|^2}{\sigma_{w}^2}$. By defining function $\xi(\nu) = \nu - \ln{\left(1+\nu\right)}$, which is a monotonically increasing function with respect to $\nu$, we can rewrite the original expression in \eqref{KL_Div} as 
    \begin{align}\label{CovertnessLevel}
        |\bm{h}_w^H(\bm{f})\, \bm{w}(\bm{f})|^2 \leq \frac{\sigma_{w}^2}{P_t}\, \xi^{-1}\left(\frac{2\epsilon^2}{L}\right) \triangleq q , 
    \end{align}
    where $\xi^{-1}(\nu)$ is the inverse function of $\xi(\nu)$. It is evident that as $|\bm{h}_w^H(\bm{f})\, \bm{w}(\bm{f})|^2$ gets larger, the probability of Willie detecting the transmission also rises. The worst-case scenario for the legitimate system occurs when the inequality in \eqref{CovertnessLevel} becomes equality.  
    We define $q = \frac{\sigma_{w}^2}{P_t}\, \xi^{-1}\left(\frac{2\epsilon^2}{L}\right)$ as the threshold for this critical state, which will be used in the subsequent derivations. We further define two areas based on this threshold: a non-covert region centered around Bob, which must be safeguarded from Willie's presence, as any intrusion into this area by Willie would result in the detection of covert communication. Outside this region lies the covert area, deemed safe for Willie's presence without risking detection. Minimizing the non-covert area around Bob is thus essential to lower the risk of detection by Willie. 
    
    \section{Transmission Strategies}
    With the antenna array deployed at Alice, we consider the following beamfocusing schemes:   
    
    \paragraph{Linear Phase Array (LPA)} First, we employ the conventional LPA approach, where all antenna elements operate at the same carrier frequency $f_c$, i.e.,
    \begin{align}
        f_n&=f_c, & n \in \{1, \dots, N\}.\label{eq:f_n_LPA}
    \end{align}
    
    \paragraph{Frequency Diverse Array (FDA)} Second, in the so-called FDA-based approach, each antenna element operates at a slightly different frequency, with small frequency increments denoted by $f_{\Delta, n}$. Thus, the sub-carrier frequency at the $n$-th antenna is given by 
     \begin{align}
        f_n & = f_c + f_{\Delta, n} , & n \in \{1, \dots, N\},
    \end{align}
     where $f_c$ refers to the central carrier frequency. We define the frequency offset vector as $\bm{f}_{\Delta} = [f_{\Delta, 1}, \dots, f_{\Delta, n}, \dots, f_{\Delta, N}]$. Three FDA-based approaches are considered as follows. 
     
     \begin{itemize}
         \item \textbf{Linear FDA:} The transmission frequency is set to linearly increase from one antenna element to the next. Taking the center of the array as the reference, the frequency increment at the $n$-th antenna is given by
    \begin{align}
        f_{\Delta, n} & = \left(n - \frac{N+1}{2}\right) F_{\Delta}, & n \in \{1, \dots, N\},
    \end{align}
    where $F_{\Delta}$ is the constant frequency shift between adjacent antennas. We assume that $F_{\Delta} \ll f_c $ due to the FDA frequency offset constraint. We note that when $F_{\Delta}=0$, linear FDA reduces to LPA. 
    \item \textbf{Random FDA:} In the random FDA scheme, different frequencies are randomly allocated to the transmit antennas. Specifically, the frequency increment allocated to the $n$-th antenna element is given by
    \begin{align}
        f_{\Delta, n} & = k_n\, F_{\Delta}, & n \in \{1, \dots, N\},
    \end{align}
    where $k_n \sim U(-\frac{N}{2}, \frac{N}{2})$ is a uniformly distributed random variable determining the frequency allocation across different antenna elements, with $F_{\Delta}$ as the frequency increment. Thus, the total system bandwidth is $N F_{\Delta}$. Note that the mathematical formulation in terms of the product of the random variable with $F_{\Delta}$ is chosen to enhance comparability with the linear FDA case. 
    \item \textbf{Optimized FDA:} Given that the FDA frequency increments impact the covert region, a possible approach is to minimize the area of the covert region through an optimal selection of frequency increments. Herein, we define the covert region’s area as a function of the frequency offset vector, $\bm{f}_{\Delta}$, and then proceed with its optimization as detailed in the following section. 
    \end{itemize} 
    \section{FDA Optimization}  
    In this section, we examine how frequency increments affect the resulting beampattern generated by the FDA. Given Bob's spatial coordinates, $(r_b, \theta_b)$, the general beampattern generated by the FDA at Bob is given by
    \begin{align}\label{BeamPattern}
        B(r, \theta; \bm{f}_{\Delta}) = \left|\frac{\beta(r)\beta(r_b)}{N}\sum\limits_{n=1}^{N} e^{j\frac{2 \pi f_n}{c}\left[r_n(r, \theta) - r_n(r_b, \theta_b)\right]}\right|^2, 
    \end{align}
    where $r_n(r_i, \theta_i)$ is defined as in \eqref{FresnelApp}. 
    According to \eqref{CovertnessLevel}, the covertness of the system is maintained when beampattern power decreases below $q$ in \eqref{CovertnessLevel}. The boundary of the covert region consists of all points $(r_w, \theta_w)$ where the beampattern power equals the threshold $q$, i.e., 
    \begin{align}\label{Boundary}
        \mathcal{B} = \{(r_w, \theta_w) | B(r_w, \theta_w; \bm{f}_{\Delta}) = q\}.
    \end{align}
    From the above, it is evident that $\mathcal{B}$ is influenced not only by the spatial position of the receiver, but also by the frequency offsets applied to the transmit antenna elements. 
    By utilizing the principles of analytic geometry, the set of boundary points of the beam pattern, i.e., $(r_w, \theta_w)$, can be modeled as an ellipsoid with its center at Bob's location $(r_b, \theta_b)$ \cite{ma2019general}. As such, the determination of the covert region is reduced to the problem of deriving the coverage area of a rotated ellipse. 
    
    Based on \eqref{BeamPattern}, $\mathcal{B}$ is determined by 
    \begin{align}\label{BeamCorrelations}
        \left|\frac{\beta(r_w)\beta(r_b)}{N}\sum\limits_{n=1}^{N} e^{j\frac{2 \pi f_n}{c}\left[r_n(r_w, \theta_w) - r_n(r_b, \theta_b)\right]}\right|^2 = q,
    \end{align}
    which captures the beam pattern power at the boundary points. This expression also reflects the correlation between two beamfocusing vectors corresponding to the locations $(r_w, \theta_w)$ and $(r_b, \theta_b)$, as defined in \eqref{CovertnessLevel}. 
    This representation shows how the correlation between the beamforming vectors at different spatial points characterizes the boundary of the covert region. 
    
    To simplify the equation in \eqref{BeamCorrelations}, we define $\Tilde{q}=\frac{q}{\beta^2(r_w)\beta^2(r_b)}$, and $z_n = \frac{2 \pi f_n}{c}\left[r_n(r_w, \theta_w) - r_n(r_b, \theta_b)\right]$. 
    As such, the boundary points of the covert region satisfy the following equation
    \begin{align}
        N^2 \Tilde{q} & = \left|\sum\limits_{n=1}^{N} e^{j z_n}\right|^2
        \overset{(a)}{=} \sum\limits_{n=1}^{N} \sum\limits_{m=1}^{N} \cos{\left(z_n - z_m\right)} ,\nonumber\\
        & \overset{(b)}{\approx} \sum\limits_{n=1}^{N} \sum\limits_{m=1}^{N} \left[1 - \frac{1}{2}\left(z_n - z_m\right)^2\right], 
    \end{align}
    where $(a)$ holds true due to Euler’s Theorem, and $(b)$ is obtained by using the second-order Taylor series approximation of the cosine function. As a result, 
    \begin{align}\label{Ellipse}
        \sum\limits_{n=1}^{N} \sum\limits_{m=1}^{N} \left(z_n - z_m\right)^2 
        = 2 N^2 (1-\Tilde{q}). 
    \end{align}
    By expanding the left-hand side of \eqref{Ellipse} and replacing $f_n = f_c + f_{\Delta, n}, \forall n$, we obtain 
    
    \begin{align}\label{Expand_Ellipse}
        \sum\limits_{n=1}^{N} \sum\limits_{m=1}^{N} & \frac{4\pi^2}{c^2} 
        \bigg[\left(f_c + f_{\Delta, n}\right)\left[r_n(r_w, \theta_w) - r_n(r_b, \theta_b)\right] \nonumber\\[-0.3cm]
        & - \left(f_c + f_{\Delta, m}\right) \left[r_m(r_w, \theta_w) - r_m(r_b, \theta_b)\right] \bigg]^2 . 
    \end{align}
    Simplifying \eqref{Expand_Ellipse} and defining $\Delta r = r_w - r_b$ and $\Delta \theta = \theta_w - \theta_b$, we can rewrite \eqref{Ellipse} as an standard elliptic equation as \cite{li2022analytical}, %(see \cite{li2022analytical}) %Appendix \ref{appendix:A})
    \begin{align}\label{Ellipse_Area}
        & g_1(\bm{f}_{\Delta}) \left(\Delta r\right)^2 + g_2(\bm{f}_{\Delta}) \Delta r\,\Delta\theta\nonumber\\ 
        & + g_3(\bm{f}_{\Delta}) \left(\Delta \theta\right)^2 -2 N^2 (1-\Tilde{q}) =0, 
    \end{align}
    where $g_1(\bm{f}_{\Delta})$, $g_2(\bm{f}_{\Delta})$, and $g_3(\bm{f}_{\Delta})$ are given as
    \begin{align}
        g_1(\bm{f}_{\Delta}) & = \frac{4\pi^2}{c^2} \sum\limits_{n=1}^{N} \sum\limits_{m=1}^{N} \left(\psi_n - \psi_m\right)^2, \\
        g_2(\bm{f}_{\Delta}) & = \frac{4\pi^2 f_c d \cos{\theta_b}}{c^2} \sum\limits_{n=1}^{N} \sum\limits_{m=1}^{N} \left[\left(\psi_n - \psi_m\right)\left(n - m\right)\right],\nonumber\\[-0.3cm]
        \\
        g_3(\bm{f}_{\Delta}) & = \frac{4\pi^2 f_c^2 d^2 \cos^2{\theta_b}}{c^2} \sum\limits_{n=1}^{N} \sum\limits_{m=1}^{N} \left(n - m\right)^2, \label{g_3}
    \end{align}
    with $\psi_n = \frac{f_c d^2 n^2}{2 r_b^2} - f_{\Delta, n}, \forall n$. By expanding the double summation in \eqref{g_3} and simplifying the expression, we obtain 
    \begin{align*}
        \sum\limits_{n=1}^{N} \sum\limits_{m=1}^{N} \left(n - m\right)^2 = N^2\left(N^2-1\right)/6,
    \end{align*}
    which is a constant that depends only on $N$. 
    Then, the beampattern area of the ellipse in \eqref{Ellipse_Area} can be obtained as 
    \begin{align}\label{S_ellipse}
        S_{\text{ellipse}}(\bm{f}_{\Delta}) = \frac{2 \pi N^2 (1-\Tilde{q})}{\sqrt{g_1(\bm{f}_{\Delta})g_3(\bm{f}_{\Delta}) - g_2^2(\bm{f}_{\Delta})}},
    \end{align}
    that shows the relationship between the FDA-based beampattern and the frequency offset vector $\bm{f}_{\Delta}$. Now, the optimal frequency offsets to minimize the covert region area, $S_{\text{ellipse}}$, can be obtained by solving the following optimization problem\vspace{-0.2cm}
    \begin{subequations}\label{eq:OptProb}
        \begin{align}
        \max_{\bm{f}_{\Delta}} \quad & g_1(\bm{f}_{\Delta})g_3(\bm{f}_{\Delta}) - g_2^2(\bm{f}_{\Delta}) \tag{\ref{eq:OptProb}}\\
        \mathrm{s.t.} \quad
            & f_{\Delta, n} \in [-F_{\Delta}/2 ,\, F_{\Delta}/2]. \label{eq:Constraint_limits}
        \end{align}
    \end{subequations}   
    The problem \eqref{eq:OptProb} is a nonlinear programming problem that can be solved using nonlinear programming solvers, which iteratively adjust the optimization variable $\bm{f}_{\Delta}$ to satisfy the constraint \eqref{eq:Constraint_limits}. The resultant solution from \eqref{eq:OptProb} is then applied to control the covert region in the optimized FDA scheme. 
    % \vspace{-0.3cm}
    \section{Numerical Results} 
    In this section, numerical results are provided to validate the effectiveness of the described near-field transmission schemes. Thereby, if not specified otherwise, Alice is equipped with a $64$-element antenna array operating at a central frequency of $f_c = \SI{3}{GHz}$, with a frequency increment of $F_{\Delta} = \SI{1}{MHz}$. Blocklength is set to $L=100$, and $\delta = 1\mathrm{e}{-5}$. Bob is located at ($r_b=\SI{7.0711}{m}$, $\theta_b=45^{\circ}$). Willie is mobile and free to move, making his exact location unknown, and he could potentially be anywhere within the area of interest. To be more specific, we assume a fine grid for the coordinates of the warden, denoted as $x_w, y_w \in [0, 40] \SI{}{m}$ with increments of $\SI{0.01}{m}$, i.e., $[x_w, y_w]$ pair corresponds to the coordinates of all possible locations of Willie. The transmit power budget is set to $P_t = \SI{20}{dBm}$, and the noise power at both Bob and Willie is set to $\sigma_{b}^2 = \sigma_{w}^2 = \SI{-60}{dBm}$. 
    \begin{figure*}
        \begin{subfigure}{0.24\linewidth}
            \includegraphics[width=\linewidth]{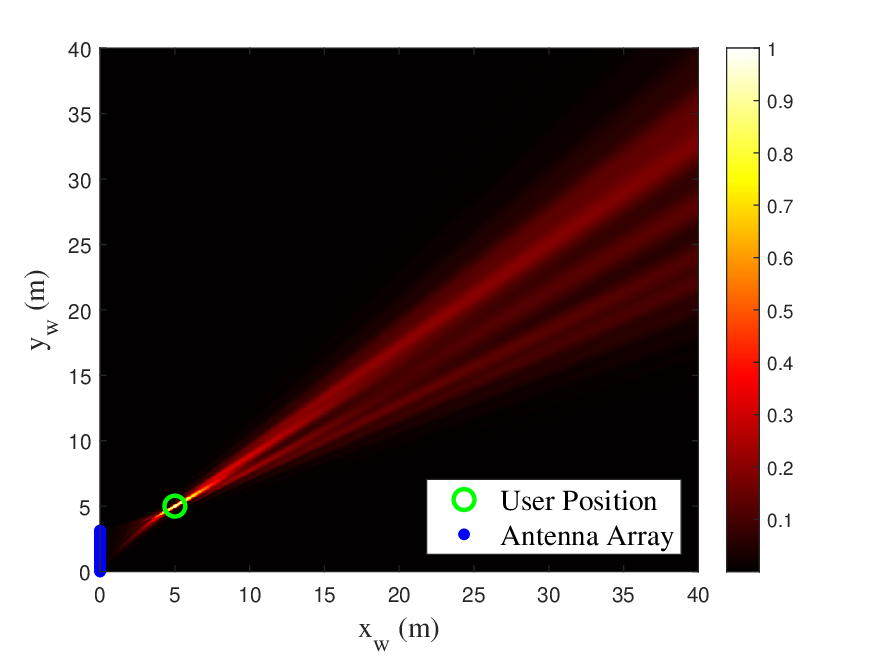}
            \caption{LPA}
            \label{fig:HeatMap_ULPA}
        \end{subfigure}
        \hfill
        \begin{subfigure}{0.24\linewidth}
            \includegraphics[width=\linewidth]{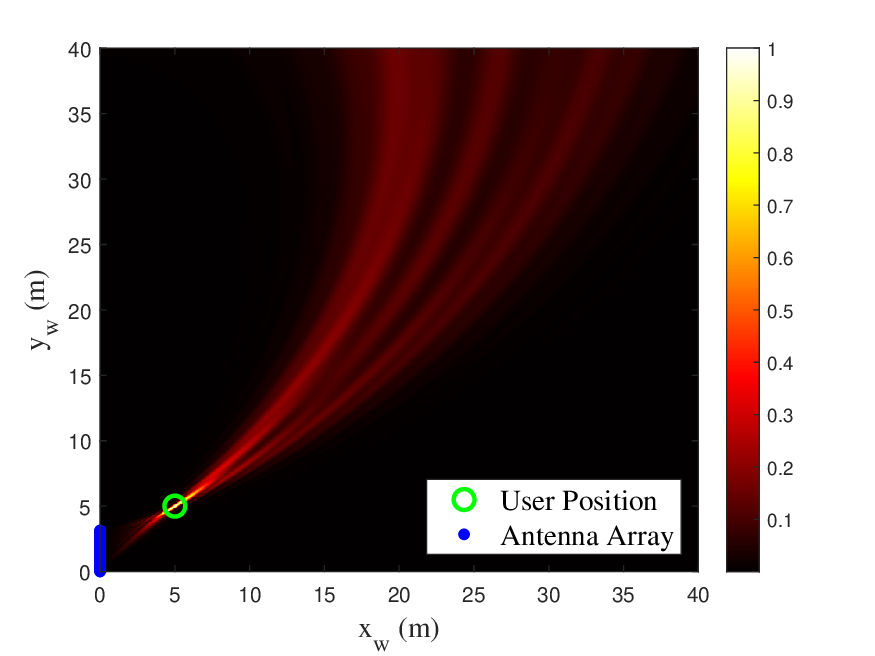}
            \caption{Linear FDA}
            \label{fig:HeatMap_LFDA}
        \end{subfigure}        
        \hfill
        \begin{subfigure}{0.24\linewidth}
            \includegraphics[width=\linewidth]{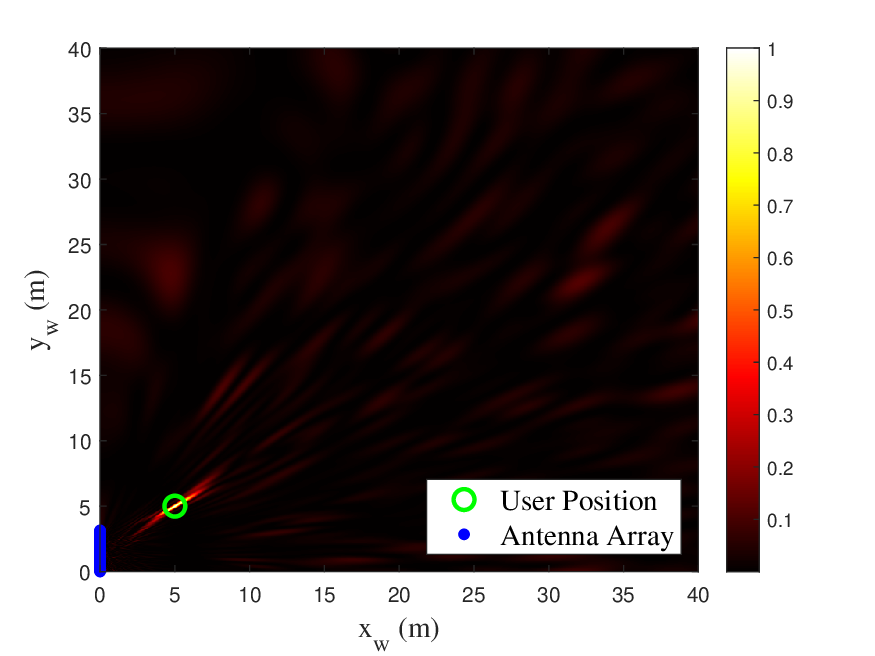}
            \caption{Random FDA}
            \label{fig:HeatMap_RFDA}
        \end{subfigure}
        \hfill
        \begin{subfigure}{0.24\linewidth}
            \includegraphics[width=\linewidth]{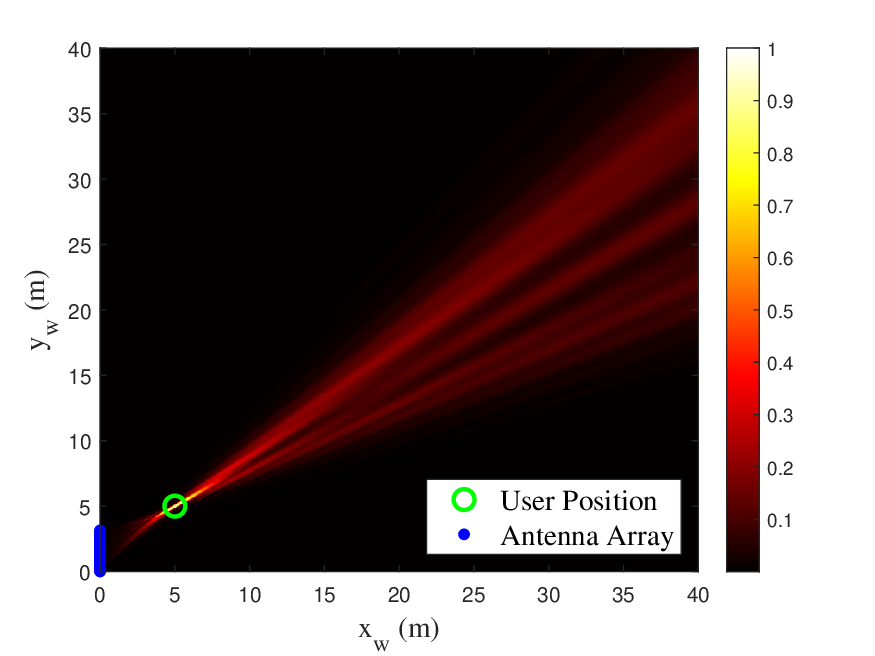}
            \caption{Optimized FDA}
            \label{fig:HeatMap_RFAS}
        \end{subfigure}  
    \caption{Normalized beampattern for different transmission strategies when Willie is positioned at $[x_w, y_w]$.}
    \label{fig:HeatMap_All}
    \end{figure*}
    \begin{figure*}
        \begin{subfigure}{0.24\linewidth}
            \includegraphics[width=\linewidth]{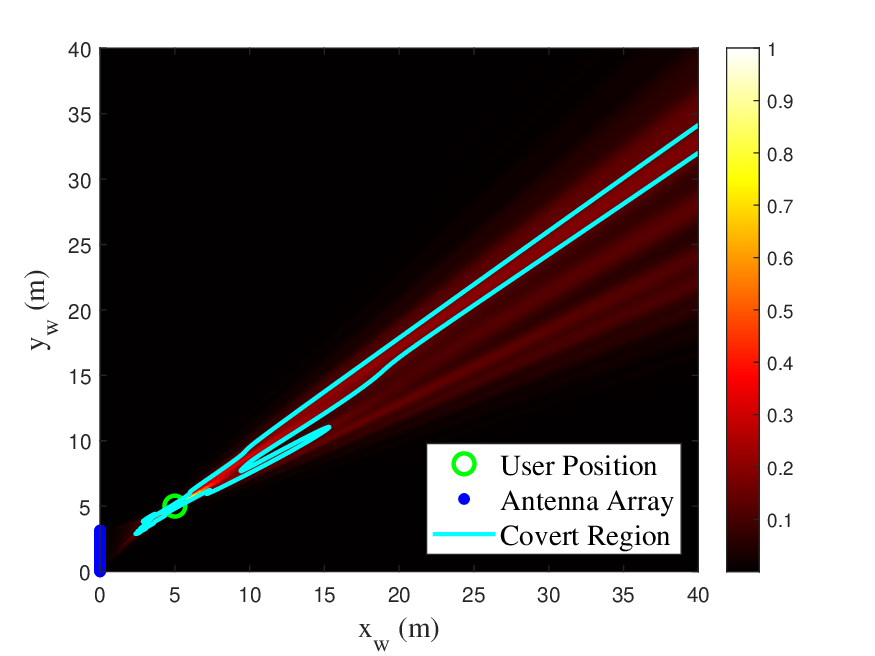}
            \caption{LPA}
            \label{fig:VulReg_ULPA}
        \end{subfigure}
        \hfill
        \begin{subfigure}{0.24\linewidth}
            \includegraphics[width=\linewidth]{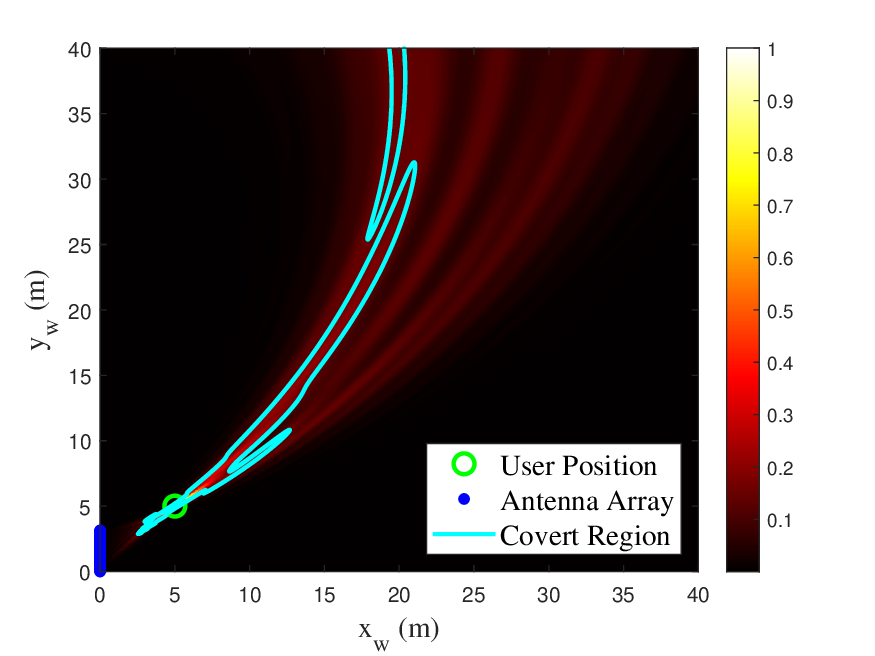}
            \caption{Linear FDA}
            \label{fig:VulReg_LFDA}
        \end{subfigure}        
        \hfill
        \begin{subfigure}{0.24\linewidth}
            \includegraphics[width=\linewidth]{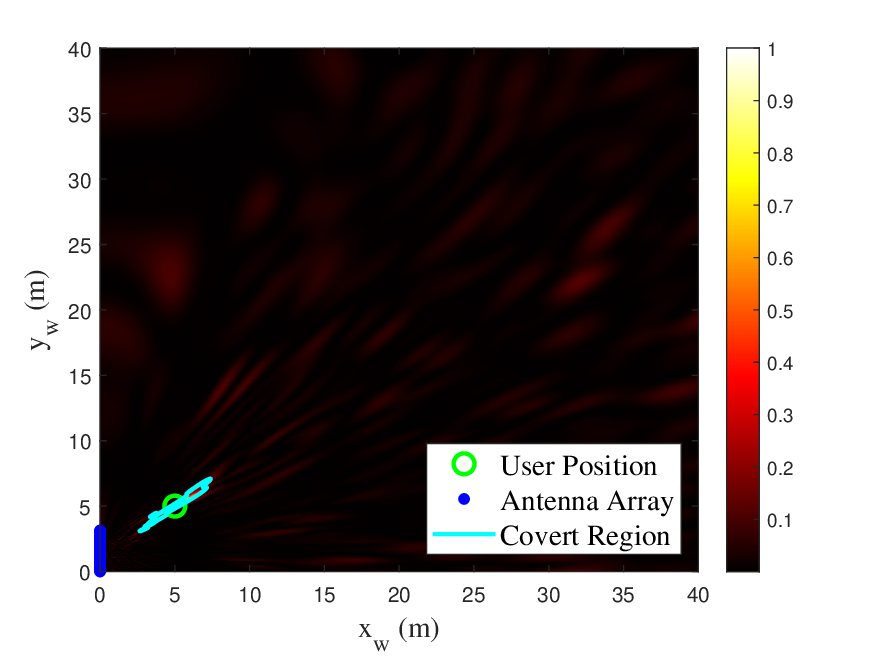}
            \caption{Random FDA}
            \label{fig:VulReg_RFDA}
        \end{subfigure}
        \hfill
        \begin{subfigure}{0.24\linewidth}
            \includegraphics[width=\linewidth]{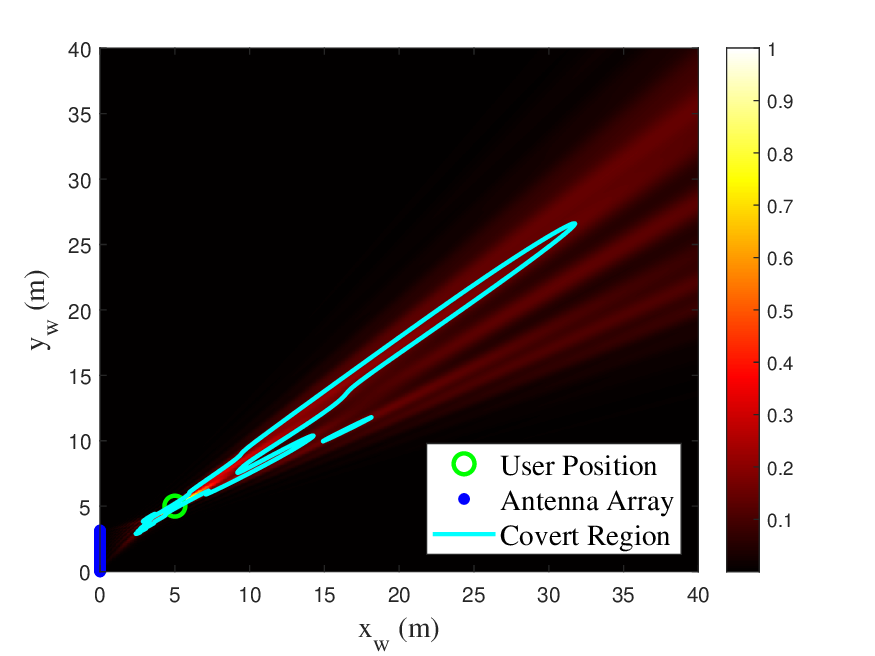}
            \caption{Optimized FDA}
            \label{fig:VulReg_RFAS}
        \end{subfigure}  
    \caption{Boundary between the non-covert area (inside the contour) and covert area (outside) for different transmission strategies.}
    \label{fig:VulReg_All}
    \end{figure*}
    Fig.~\ref{fig:HeatMap_All} illustrates the beampattern observed under the different schemes for all possible locations of Willie. The plots show that the area of the non-covert region (determined by the red color) shrinks when applying different transmission schemes, as they reduce the beampattern power at locations other than Bob's. This is achieved by minimizing the channel similarity between the Alice-Bob and Alice-Willie channels across all possible locations of Willie.  
    To visually compare the non-covert region across all schemes, we set a threshold of \SI{10}{\%} of Bob's received energy (by adjusting the parameters $\sigma_{w}^2, P_t, \epsilon, L$) to define the non-covert region. The share of the area with energy levels below this threshold are highlighted in Fig.~\ref{fig:VulReg_All}, illustrating the extent of the covert region under different transmission strategies. 
    
\begin{figure*}
        \begin{subfigure}{0.49\linewidth}
            \includegraphics[width=\linewidth]{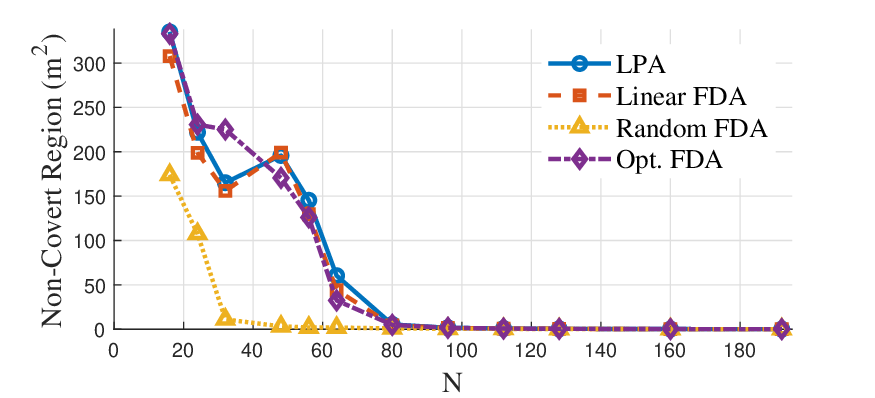}
            \caption{$F_{\Delta} = \SI{1}{MHz}$}
            \label{fig:Array_Numbers}
        \end{subfigure}
        \hfill
        \begin{subfigure}{0.49\linewidth}
            \includegraphics[width=\linewidth]{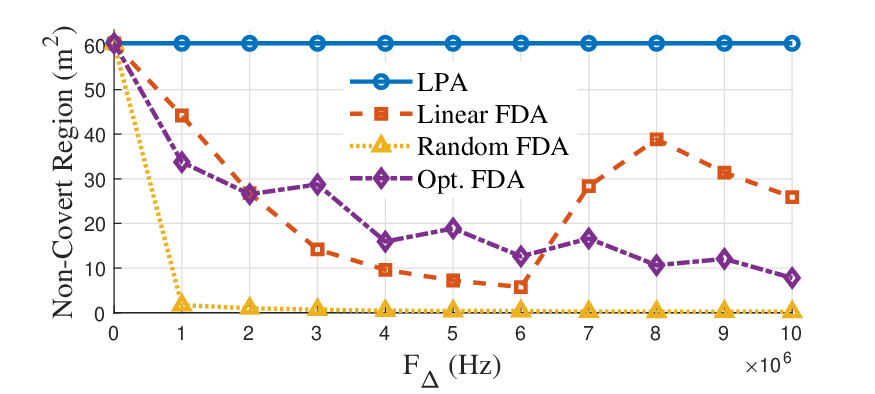}
            \caption{$N=64$}
            \label{fig:Freq_Increments}
        \end{subfigure}
            \caption{Area of the non-covert regions at the cell area for different transmission strategies.} 
            \label{fig:Region_Percentage}\vspace{-0.4cm}
    \end{figure*}

    Fig.~\ref{fig:Array_Numbers} and Fig.~\ref{fig:Freq_Increments} illustrate the area of the non-covert regions within the area of interest for different transmission strategies versus antenna number $N$ and frequency increment $F_{\Delta}$, respectively. As Fig.~\ref{fig:Array_Numbers} shows, more antennas significantly reduce the non-covert region across all schemes. This is because a larger number of antennas enables more precise beamfocusing, thus concentrating the signal energy toward Bob and minimizing leakage toward other regions. Random FDA performs particularly well in reducing the non-covert region.  
    From Fig.~\ref{fig:Freq_Increments}, it can be seen that LPA is unaffected by the parameter $F_{\Delta}$ due to \eqref{eq:f_n_LPA}, while all other schemes benefit from larger $F_{\Delta}$. Random FDA shows the largest improvement in reducing the non-covert region as $F_{\Delta}$ increases. This is due to the randomization of frequency across antenna elements, allowing for better spatial control of the beampattern.    

\begin{figure*}
        \begin{subfigure}{0.49\linewidth}
            \includegraphics[width=\linewidth]{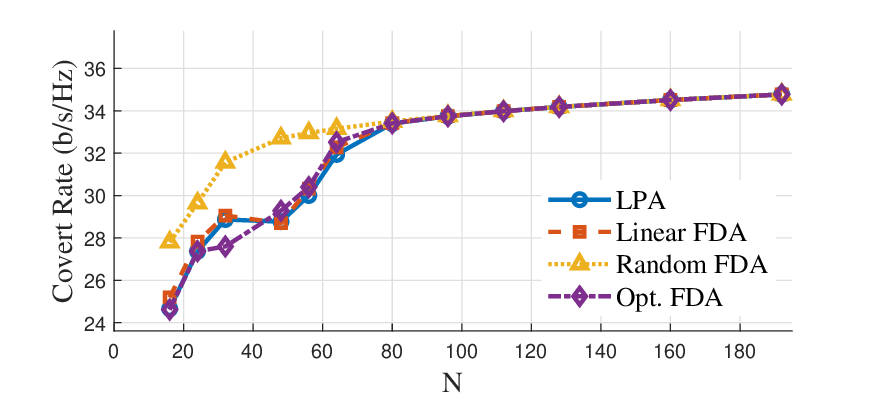}
            \caption{$F_{\Delta} = \SI{1}{MHz}$}
            \label{fig:Rate_vs_ArrayNumbers}
        \end{subfigure}
        \hfill
        \begin{subfigure}{0.49\linewidth}
            \includegraphics[width=\linewidth]{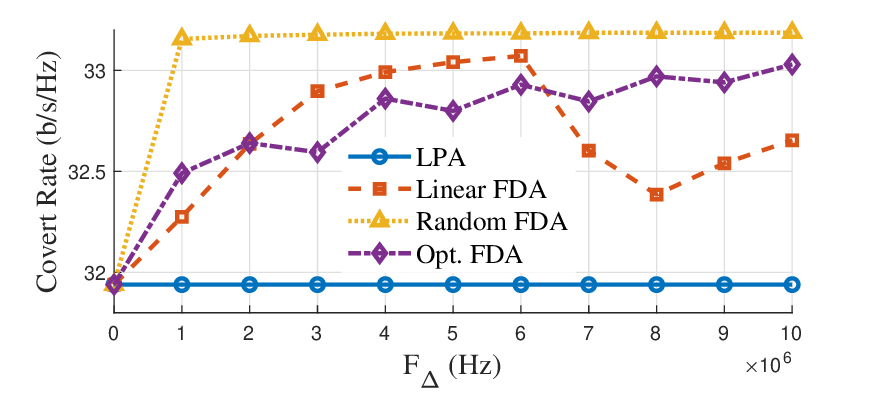}
            \caption{$N=64$}   
            \label{fig:Rate_vs_FreqIncrements}
        \end{subfigure}
            \caption{Covert rate for different transmission strategies and parameter choices.} % (a) $F_{\Delta} = \SI{1}{MHz}$. (b) $N=64$. }
            \label{fig:CovertRates}\vspace{-0.4cm}
    \end{figure*}

    Fig.~\ref{fig:CovertRates} illustrates the covert rate versus $N$ and $F_{\Delta}$, respectively. The plots show the average achievable rate, calculated with random locations of Willie. Specifically, we set $R_b=0$ if Willie is located in the non-covert area and apply \eqref{CovertRate} otherwise. As we increase $N$ in Fig.~\ref{fig:Rate_vs_ArrayNumbers}, the covert rate for all schemes generally increases. This was expected because more antennas lead to better beamfocusing gain and thus a higher rate toward Bob. In Fig.~\ref{fig:Rate_vs_FreqIncrements}, the covert rate is shown as a function of $F_{\Delta}$, with random FDA consistently outperforming the other schemes. When comparing this to Fig.~\ref{fig:Freq_Increments}, we observe an opposite trend, which aligns with the analysis. This inverse relationship arises because a smaller non-covert region results in a higher covert rate.

\section{Conclusion} 
    In this paper, we have investigated a non-covert region surrounding legitimate devices. Since near-field beampatterns are inherently distance-angle-dependent, we have employed various FDA-based transmission strategies, incorporating random carrier frequencies at the transmitter to manipulate the two-dimensional beampatterns. The numerical results demonstrate that the proposed schemes are effective in reducing the area that needs protection from potential warden detection. This effectively expands the covert region, allowing the warden to move freely within a broader area without compromising system covertness. More specifically, the proposed strategies generate more focused beampatterns directed at the legitimate user, while simultaneously reducing energy leakage toward unintended receivers, considerably enhancing the system’s covertness. Additionally, the non-covert region shrinks as the number of array elements and the frequency increments at the transmitter increase. A larger antenna array and wider frequency intervals contribute to more focused beams, further improving covertness. 

\bibliographystyle{IEEEtran}


\end{document}